\documentclass[ras]{agu2001}
\usepackage{graphicx}
\authorrunninghead{Briggs \& Kocz}

\titlerunninghead{Overview of RFI mitigation techniques}

\authoraddr{F.H. Briggs,
Research School of Astronomy \& Astrophysics,
Mount Stromlo Observatory,
Cotter Road, Weston, ACT 2611, Australia
(fbriggs@mso.anu.edu.au)}

\begin{document}

%
%
%
%
%

%
%

\title{Overview of Technical Approaches to RFI Mitigation}
%

%
%


\author{F. H. Briggs}
\affil{Research School of Astronomy and Astrophysics, The Australian National
University, Canberra, ACT, Australia, and Australia Telescope National Facility,
CSIRO, Epping, NSW, Austrlia}

\author{J. Kocz}
\affil{Department of Engineering, The Australian National
University, Canberra, ACT, Australia}

\begin{abstract}
This overview provides an interface between lines of thought on rfi
mitigation in the fields of radio astronomy and signal processing.
The goal is to explore the commonality
of different approaches to help  researchers in both fields
interpret each others' concepts and jargon. The paper elaborates on
the astronomers' concept of gain closure relations and how they may be used
in a ``self-calibrating'' system of rfi cancellation. Further discussion of
the eigen decomposition method in terms of rfi power and antenna gains
introduces adaptive nulling and rfi cancellation through matrix partitioning.
While multipath scattering appears at first glance to be fatal to methods of
rfi cancellation, its effect is easily incorporated in frequency dependent gain
coefficients under many circumstances.
\end{abstract}

%
%

%

\begin{article}

%
%

\section{Introduction}

Astronomers are designing the next generation radio telescopes with
the capabilities to solve many current puzzles about the origin
of the planets, stars, galaxies and Universe itself. Since the telescope
designs are
one to two orders of magnitude more powerful than the best present
telescopes, astronomers know that the new instruments are capable of
making discoveries that will open whole new fields of research.
New telescopes will not just solve old problems, but will certainly
do unexpected things that have been impossible in the past.

The same advances in technology that make new radio astronomy possible
are closely linked to the advances in telecommunications that are
beginning to saturate the radio spectrum.  Any new radio telescope
will have additional layers of signal processing built in from the
design stages that are intended to enable it to see through the
background radio interference as well as possible.  In this regard,
modern technical solutions offer great hope, but in reality they
are complex and expensive, and the resources  (in terms of time and money)
will probably not exist  to allow radio astronomers to apply these
methods effectively enough to
keep up with the rising earth-generated radio background unless there
is active coordination with the telecommunications industry.

Communications signals are fully polarized, coherent emission from
carefully modulated currents in the broadcasting antennas;
a communications signal is broadcast to provide maximum efficiency in use
of a limited portion of the radio spectrum defined by its
allocated bandwidth $BW$. The astronomical
signals originate mainly in the incoherent emissions from huge clouds
of faintly emitting atoms, molecules or energetic electrons whose
``astronomical intensity'' is a collective effect arising from their
large numbers. The radiated powers are added incoherently, except in
a few cases, such as the coherent addition of radiated electric
field in the ionospheres of pulsars, Jupiter's ionosphere,
the active regions of the Sun or celestial masers. 
Astronomical signals in general
are well represented by stochastic or gaussian noise.

Telecommunication reception requires decoding the fluctuations in the
electric field of the signal, so that the field at each time step $T$
(where $T\approx 1/BW$) reliably carries information. The information that
astronomers extract from their signals has a more statistical character,
such as total intensity of the power per unit bandwidth (called
{\it flux density}) of an astronomical
source, along with other more detailed statistical measures such as the
net polarization of the signal power and the variation of the flux
density as functions of frequency and time.  For the large variety of
celestial sources that fall in the class called `continuum sources,'
which radiate a smooth spectrum over a wide range of frequencies,
more power can be received from the source by simply increasing the
bandwidth of the observation.

Astronomers have not generally been interested in measuring the
electric field strength as a function of time, although this is likely to
change for some astronomical applications, such as rapidly time-variable
phenomena (for example: pulsars, transient bursts, signals from energetic particles
impinging on the upper atmosphere \cite{hueg03,supr03}), 
especially with the  more
sensitive telescopes being planned for the future.
The need for RFI mitigation has
also increased astronomers' awareness of the need for processing their E-field
signals in real-time at full bandwidth, which is much more demanding and
expensive than their historical procedures (\cite{barn98}).

While there do exist frequency bands that have been reserved for astronomy, 
receivers have become so sensitive
that low level spurious harmonics and weak leakage from communication and
navigation services are strong enough
to corrupt and confuse the astronomical signals. Furthermore,
pursuing answers to scientific questions drives astronomers to observe outside of the
reserved bands. For example, neutral atomic hydrogen has a well known radiative 
transition at rest frequency $f_o$$\:=\:$$1420.4057$~MHz, for which there is a reserved astronomy
band. However, astronomers now wish to trace the evolution of hydrogen in galaxies 
over cosmic time, and since observing galaxies at redshift $z\approx 1$ would
correspond to looking back in time to when the universe was less than half its current
age, radio astronomers pursue the hydrogen line to frequencies of 
$f=f_o/(1+z)\approx 700$~MHz or lower, where the radio spectrum is already in heavy
use.

Astronomers have had great success in achieving high angular resolution and high dynamic
range imaging. The question naturally arises, ``why not simply image the rfi sources
along with the celestial ones?''  Under some circumstances, this is a successful
approach (e.g. \cite{perl04}); however, the character of many RFI signals -- variability,
non-gaussianity -- may not be compatible with the necessary conditions for obtaining
sufficiently high-dynamic range aperture synthesis.

As in all of radio science, improvements in instrumentation and signal processing
are
providing the tools for performing astronomical observations in radio
rich environments. These include:

\begin{enumerate}

\item{ {\bf Robust receivers} Modern receivers benefit from linearity over a 
large dynamic range to  avoid generation of spurious signals
  through intermodulation within the receiving system.
  Since astronomical signals are noise-like, astronomers have historically
 processed their
  signals with custom built correlators that operate at 1 or 2 bit precision,
  at the cost of adding a modest percentage of quantization noise.
  Newer telescopes that will operate in populated areas are designed to
  process data with 10 to 14 bits of precision, a cost only now affordable
  due to advances in low-cost digital processing electronics.}

\item{ {\bf Sophisticated signal recognition and blanking mechanisms}. 
Historically, this was performed in the post-detection domain, where it
was labor intensive and called ``data editing.''
If done early in the receiving system chain, blanking can ease the large dynamic range
requirements of the receiver backend  (reducing the required bits of precision
for example).  It is always a concern that a process so violently non-linear
as blanking may preclude the application of some classes of rfi cancellation
for which linear systems are a prerequisite. These topics in rfi mitigationnhave
been reviewed by Fridman and Baan (\cite{frid01}) and are addressed by several papers
at this conference.}

\item{{\bf RFI cancellation} A number of recent papers have begun to explore methods of
rfi cancellation and subtraction(\cite{barn98,elli00,lesh00,brig00}). The
series of papers by Leshem, van der Veen, Boonstra and colleagues
have been especially important in applying the techniques of signal processing in the
radio astronomical context (see \cite{boon04} in these proceedings for complete 
references).}

\end{enumerate}

The bulk of this paper forms a mini-tutorial
covering several issues in rfi cancellation
that were sources of confusion to the authors and their colleagues.
The goal is to ease the transfer of technology between the astronomical
and signal processing communities.

Of course, the best starting place for radio astronomy continues 
to be siting telescopes
in radio
quiet environments, and astronomy will always rely on good will and cooperation
with radio-active spectrum users for communication and navigation services.

%
%


%
%

\section{Radio Astronomical Arrays} 

\begin{figure*}
\noindent\hspace{1.2cm}\includegraphics[width=33pc]{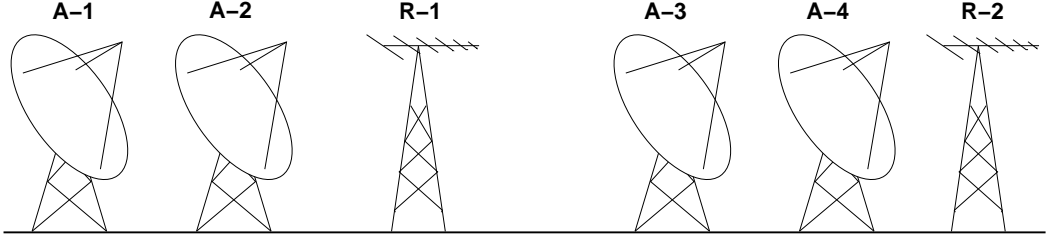}
\caption{Astronomical telescope arrays with (1) conventional dish antennas
pointed toward celestial radio sources
 and (2) reference antennas to provide increased gain in  the
direction of known interfering signals.
}\label{antenna_schematic.fig}
\end{figure*}

Modern radio telescopes operate in arrays in order to economically obtain
sensitivity and angular resolution through the combination of the signal
from the elements of the array (Figure~\ref{antenna_schematic.fig}). 
The radio astronomical convention is to
denote antenna separations by a vector quantity $\overrightarrow{B_{ij}}$, indicating
the relative positions of the antenna phase centers for antennas $i$
and $j$. These ``interferometer baselines'' are
conveniently measured in wavelengths $\overrightarrow{B_{\lambda}}$ for which the
effective angular resolution of each interferometer pair in the
array is $\theta_{ij} = 1/B_{\lambda,ij}$ radians.
Definition of a cartesian coordinate system ($u$-$v$-$w$) with its
$w$ axis pointing in the direction of a radio source allows $\overrightarrow{B_{\lambda}}$
to be specified by its components $u=B_{\lambda,u}$, $v=B_{\lambda,v}$,
and $w=B_{\lambda,w}$. An array of elements at fixed locations will experience
systematic variation of $u$, $v$ and $w$ as the Earth rotates.
The $w$ component specifies the relative propagation
delay between two elements of the array $t_{ij}=w_{ij}\lambda/c$, while the
$u$ and $v$ components appear in the key equation of aperture 
synthesis,

\begin{eqnarray}
V(u,v)= \frac{2k}{\lambda^2}\int\hspace{-.1cm} p_B(x,y)T_b(x,y) e^{i2\pi(ux+vy)} dx\:dy
\label{vis.eqn}
\end{eqnarray}
relating the ``fringe visibility''
$V(u_{ij},v_{ij})$ sensed by each interferometer pair $i,j$ in response
to the
angular brightness temperature distribution of the sky $T_b(x,y)$
modulated by $p_B(x,y)$, the primary beam gain function of the array
elements. The
strength and phase of the complex visibility result from the
fourier integral of $p_B(x,y)T_b(x,y)$ over the celestial source.
Here, $k$ is the Boltzmann constant, and $x$ and $y$ are angular
coordinates measured in the plane of the sky.
While this integral
and its inversion form the essence of the aperture synthesis technique,
a full description of the method requires entire books
(e.g. \cite{thom86,vla1999}).

Interferometer arrays measure a $V_m(u,v)$ for each baseline through
the cross correlation $C_{ij}$ of the signals $s_i$ and $s_j$ from two elements:
\begin{eqnarray}\hspace{1.5cm}
V_m(u_{ij},v_{ij}) & = & <g_is_ig_j^*s_j^*>\\
&=&C_{ij}
\nonumber
\label{obsvis.eqn}
\end{eqnarray}
The signals entering the correlator have been
scaled by the complex gain factors $g_i$ that quantify the
differences in the signal paths from one array element to the next,
and $*$ indicates complex conjugation. In general,
$V_m(u,v,f)$, $g_i(f)$ and $s_i(f)$ are functions of frequency, and
$C_{ij}(f)$ is the observed cross power spectrum.
Provided the gains are
constant over the integration time of the cross correlation product in
Eqn.~\ref{obsvis.eqn}, the expression becomes 
$V_{ij}=V_m(u_{ij},v_{ij})= g_ig_j^*<s_is_j^*>$.

\section{Gain closure}

The concept of ``closure relations'' becomes clear when combinations of
observed correlation products are formed:
\begin{eqnarray}\hspace{0.2cm}
Q = \frac{V_{ij}V_{kn}}{V_{in}V_{kj}}&=&\frac{g_ig_j^*<s_is_j^*>g_kg_n^*<s_ks_n^*>}
{g_ig_n^*<s_is_n^*>g_kg_j^*<s_ks_j^*>}\nonumber\\
&=&\frac{C_{ij}C_{kn}}{C_{in}C_{kj}}
\end{eqnarray}
The quantity $Q$ is a ratio of true visibilities for the radio source,
regardless of whether the individual complex gains $g_i$ are known accurately or
not.  

The special case of an unresolved point source is of interest, since
substituting a $\delta$-function for $T_b(0,0)$ in Eqn.~\ref{vis.eqn} yields a
visibility function that is constant for all $u,v$, implying that $Q=1$ and the
$<s_is_j^*>$ are equal for all $i,j$, so that the right
hand side of the equation reduces to an identity relation among the gain 
coefficients.

The gain closure relations form the motivation for approaches to
rfi mitigation since rfi sources can be treated as point sources and
the cross correlations between rfi reference antennas and the astronomical
signals provide the necessary information for computing the rfi contamination
without precise knowledge of the gains of the element sidelobes through which 
the rfi enters the telescopes.  These methods are in essence ``self-calibrating.''


%
%

\section{Cancellation}

Most rfi cancellation schemes involve first identifying the interferer, followed
by deducing the strength of its contamination within the astronomical data stream,
and finally, precise subtraction of interferer's contribution.

\begin{figure}
\noindent\includegraphics[angle=-90,width=20pc]{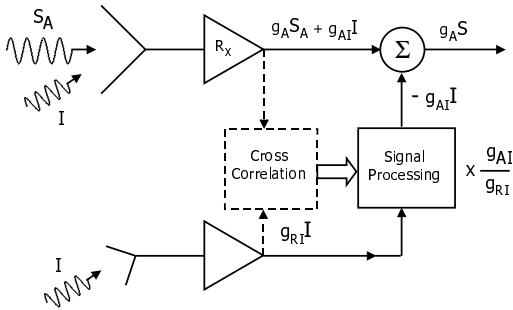}
\caption{Example of interference cancellation scheme. In this case, there
are two signal paths: (1) the astronomical signal $S_A$ entering the main antenna but
corrupted by interference $I$, and (2) a reference antenna to pick up the interfering
signal apart from $S$. The goal is for the signal processing to obtain a suitable signal
for subtraction from the astronomical signal path to leave only an undistorted $S$.
Techniques discussed here use cross correlation between Paths A and R to determine
the filter coefficients.
}\label{canc_schematic.fig}
\end{figure}

Fig.~\ref{canc_schematic.fig} illustrates one example of how that might be
accomplished using the signal $I$ from an rfi reference antenna that is cross correlated
with the astronomical signal $S_A$ in order to deduce the inter-coupling of the two
signal paths.  Several variations of implementing this scheme through the
gain closure relations of cross power spectra 
were explored by \cite{brig00}. Other schemes are implemented in the time
domain (e.g. \cite{barn98,kest04}).

Separate rfi reference antennas serve two purposes: (1) They improve the signal-to-noise
ratio on the interfering signal, in order to minimize the addition of
noise to the astronomical observation
during the application of rfi subtraction, and (2) with increased gain toward the
interferer but decreased gain toward the celestial source, they avoid inadvertent
subtraction of the astronomical signal along with the interference. It is important that
the interference-to-noise ratio be better in the reference channel than the astronomy
channels (\cite{brig00}), but this is generally true since rfi enters the astronomy
telescopes through weak sidelobes (roughly at the level of an 
isotropic antenna $\sim$0~dBi), while simple reference antennas can deliver 10~dB or
more gain).

The gain information contained in the cross power spectrum $C_{AR}(f)$
 between $(g_AS_A+g_{AI}I)$ and $g_{RI}I$ allows
the rfi contamination of the voltage spectrum $g_{AI}I(f)$ in the
astronomical data stream to be estimated from the reference voltage
spectrum $g_{RI}I$:

\begin{eqnarray}\hspace{1.5cm}
g_{AI}I &=& \frac{C_{AR}}{g_{RI}^*I^*}
        = \frac{g_{AI}g_{RI}^*II^*}{g_{RI}^*I^*}
\end{eqnarray}

Astronomers commonly measure the power spectrum of the signal from
the telescope $P_A(f)$, which will have both power from the celestial source
and rfi components $P_A(f)= |g_AS_A|^2+|g_{AI}I|^2$.  In this case, an
appropriate correction is preferably derived from two reference antennas
labeled R1 and R2:
\begin{eqnarray}
\label{totpwrcorr.eqn}\hspace{1.3cm}
|g_{AI}I|^2 &=& \frac{C_{AR1}C_{AR2}^*}{C_{R1R2}^*}\\
        &=& \frac{g_{AI}^{ }g_{R1}^*|I|^2g_{AI}^*g_{R2}^{ }|I|^2}{g_{R1}^*g_{R2}^{}|I|^2}
\nonumber
\end{eqnarray}
The preference for dual reference signals arises from the nature of the
noise in the cross power spectrum in the denominator of Eqn.~\ref{totpwrcorr.eqn};
the noise term $<$$N_{R1}N_{R2}^*$$>$$(f)$ is a complex cross power spectrum that averages
toward zero as integration time increases, unlike the noise $<$$|N_{R1}|^2$$>$$(f)$, which
is real and positive and which would bias the result of Eqn.~\ref{totpwrcorr.eqn}
if it were based on a single reference signal. When the subtraction is performed
after the calculation of power spectra in this way, it is often described as
``post-correlation'' cancellation.

Similar estimation for the contamination of the $V_m(u,v)$ measurement of an
interferometer pair $i,j$ of an array (Fig.~\ref{antenna_schematic.fig}) is
\begin{eqnarray}\hspace{1.0cm}
\label{crosspwrcorr.eqn}
g_{AiI}g_{AjI}|I|^2 &=& \frac{C_{AiR1}C_{AjR2}^*}{C_{R1R2}^*}\\
        &=& \frac{g_{AiI}^{ }g_{R1}^*|I|^2g_{AjI}^*g_{R2}^{ }|I|^2}{g_{R1}^*g_{R2}^{}|I|^2}
\nonumber
\end{eqnarray}
with dual reference antennae.

\begin{figure}
\noindent\includegraphics[width=19pc]{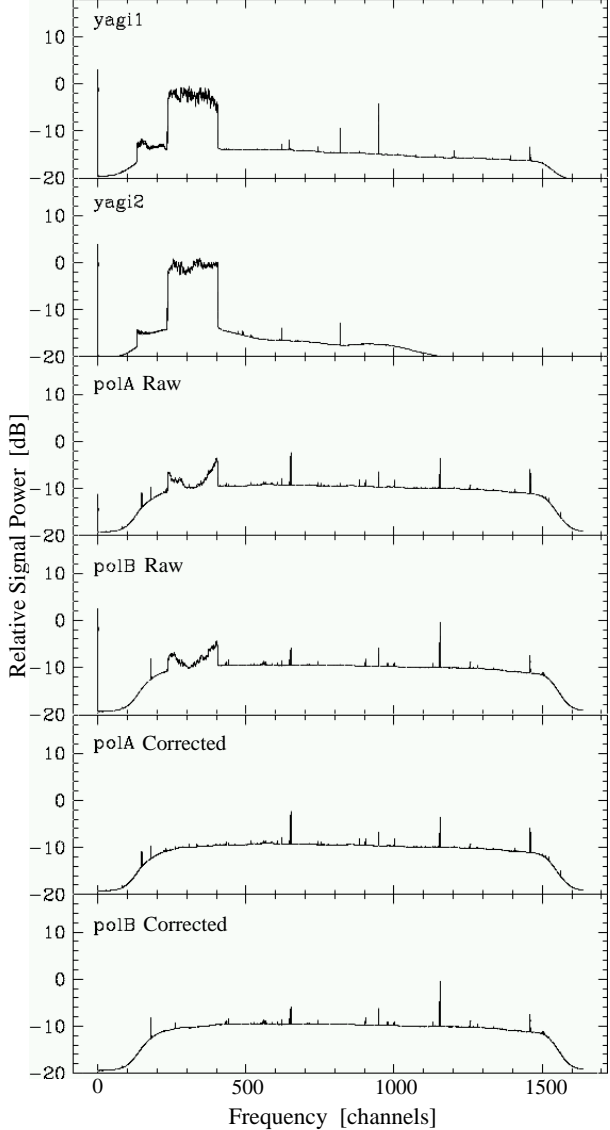}
\caption{Application of post-correlation gain-closure cancellation. Observations
in the 50cm digital TV band from the Parkes Telescope using the CPSR2 recording system.
The 64 MHz wide bands were analyzed in 16384 channel spectra.
Top panels show reference signals from two yagi reference antennas aimed at the
broadcasting tower. The astronomy total power spectra for the two polarizations of the
Parkes Telescope are shown as 1) uncorrected (central panels) and 2) corrected (lower
pannels).
}\label{six_chan.fig}
\end{figure}

Figure~\ref{six_chan.fig} gives an example of applying the relations in
Eqn.~\ref{totpwrcorr.eqn}. These Parkes observations covered a 64~MHz band
centered on 676 MHz with 16384 spectral channels (3.9~kHz resolution). 
The goal of the program is the observation of neutral hydrogen clouds 
at reshift $z\sim 1$ in absorption against high $z$ radio galaxies.
Two yagi
reference antennas were oriented to point at the broadcasting tower for digital
TV station channel 46 (652--659~MHz). The
cancellation algorithm was applied to $\sim$0.1~sec integrations, followed by
averaging for the full 16~sec scan. After cancellation of the digital TV, the noise
level and weaker rfi contamination remains at the same level as the surrounding frequency
bands; only those interferers in the yagi reception pattern have been canceled.
(\cite{kest04} give more details on the nature of the digital TV interference
in their discussion of a real-time adaptive canceler.)

\begin{figure*}
\noindent\hspace{0.3cm}\includegraphics[width=36pc]{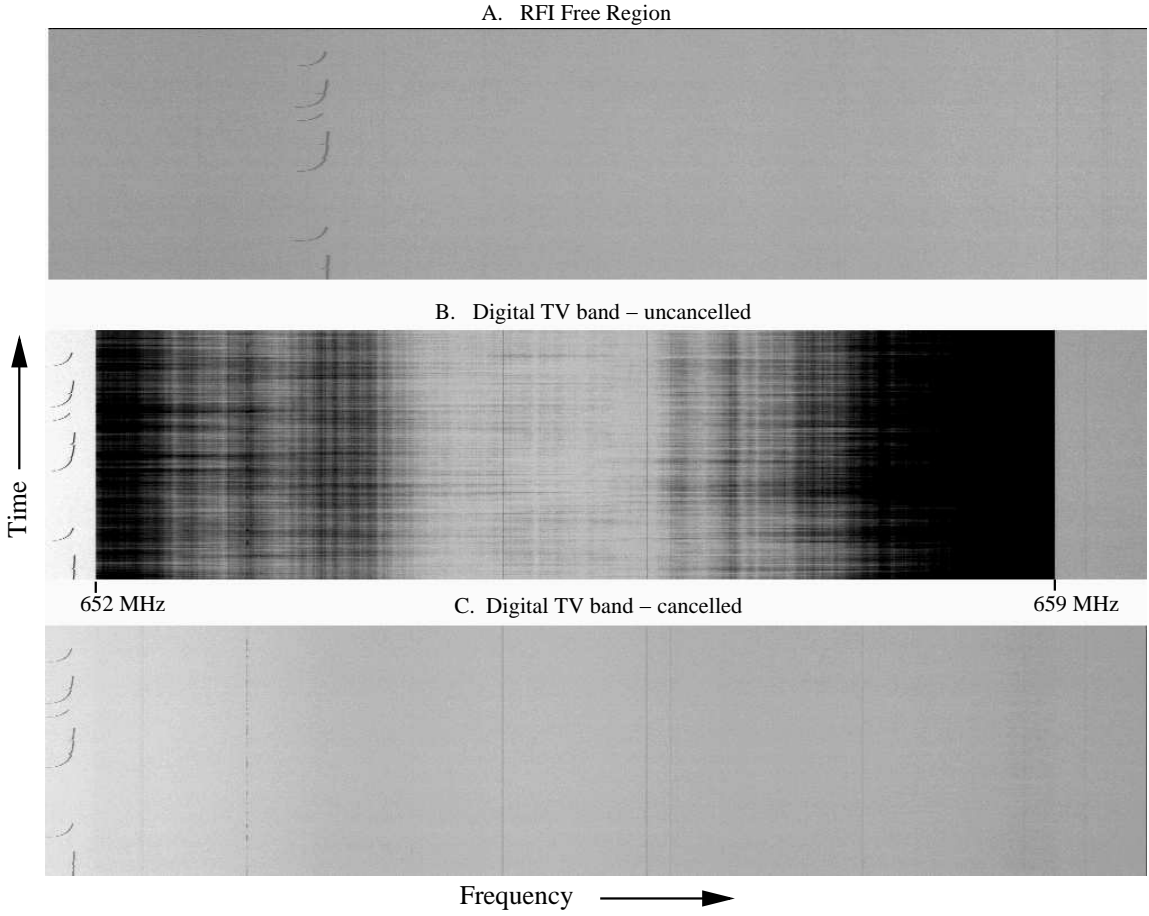}
\caption{Dynamic spectra for A) a spectral region free of digital TV contamination,
B) the spectral band containing TV channel 46, prior to cancellation, and C) the
channel 46 band after cancellation.
}\label{image.fig}
\end{figure*}

Figure~\ref{image.fig} provides a closer look at the band containing the digital TV
as a function of time through the 16 second scan -- both before and after cancellation
of the rfi signal. The noise level should be compared to that in a relatively ``clean''
band away from the TV (in the top panel of the figure). 
The clean band also contains an image of the peculiar drifting
interferer, one of whose harmonics appears to the left of the TV band in panels B and C.
The weak rfi spikes that remain throughout the spectrum (1) are narrower than the
expected HI signals, (2) appear as emission features rather than absorption lines,
and (3) take up a minor fraction of the spectrum as evidenced by the broad
expanses between the narrow lines. This is encouraging for the success of redshifted
HI studies.

\section{Multipath delays}

At first glance, multipath propagation or scattering of the interfering
signal on the way from the transmitter to either the astronomical or reference
antennas would appear to corrupt the interfering signal -- in different ways 
along the paths to the different receiving systems -- rendering it impossible to
extract an appropriate reference signal.  Fortunately, multipath is not a problem
provided the paths are stable on sufficiently long time scales.

This can be easily seen by noting that the interfering signal received by an
antenna $i_r(t)$ is the sum of different versions of the signal $i_o(t)$ broadcast
by the transmitter at the source:

$$
i_r(t) = \sum_{\alpha} h_{\alpha}(t)\star i_o(t-\tau_{\alpha})
$$
Here $h_{\alpha}(t)$ is the impulse response experienced by the interfering
signal along the path labeled $\alpha$, and $\star$ is the convolution operator.
We have chosen to enter the relative time delays of the signal explicitly in the
second term in the sum. In the frequency domain, the received spectrum $I_r(f)$
takes the form

\begin{eqnarray}\hspace{1.08cm}
I_r(f) &=& \sum_{\alpha} G_{\alpha}(f)\:I_o(f)\: e^{-i2\pi\tau_{\alpha}f}
\nonumber\\
&=& I_o(f)\sum_{\alpha} G_{\alpha}(f)\: e^{-i2\pi\tau_{\alpha}f}
\nonumber\\
&=&I_o(f)\:G_{tot}(f)
\label{phasewind.eqn}
\end{eqnarray}
where all the ugly details of the multipathing are contained in the single complex
gain
$G_{tot}(f)=\sum G_{\alpha}(f)\: e^{-i2\pi\tau_{\alpha}f}$, and the $I_o(f)$
stands cleanly apart. $G_{tot}(f)$ is likely to have strong frequency dependence,
caused by differential phase winding from the relative delays. Clearly, application
of algorithms such as Eqns~\ref{totpwrcorr.eqn} and \ref{crosspwrcorr.eqn} will
require spectral resolution for the frequency channels of $\Delta f < 1/\tau_{max}$,
where $\tau_{max}$ is the greatest relative delay. The equivalent statement for algorithms
that operate in the time domain is that the number of lags used in constructing corrective
filters must be sufficiently great to capture the full range of delays.

\section{Eigen Decomposition and Gain Calibration}

Interferometer arrays measure correlations between the signals arriving at
the array elements. Correlation matrices, such
as shown in Fig.~\ref{cross_corr_spectra.fig}, can be used to summarize the
covariances of the signals  throughout each integration time.
Since astronomers seek to detect sources that are too weak to detect in
the short time of a correlator dump interval, they extract information from the
measured covariances (or fringe visibilities) by searching for subtle correlations
through inversion of the fourier integral of Eqn.~\ref{vis.eqn}.
This is an effective way of reorganizing the correlation information by restricting
the range of interest to only those signals that could have originated from within
the field of view of their telescopes' primary beams $p_B(x,y)$. Since the Earth's
rotation causes the $u$,$v$ values to change over time, the coefficients
of the covariance matrix are expected to change with time; making maps is an
efficient way to recover the correlations in a physically meaningful way.

There are other powerful
mathematical tools for recovering information from correlation matrices,
although these are geared for studying correlations over the short integration times
for which the coefficients are expected to remain constant.
For example,
Leshem, van der Veen and Boonstra (\cite{lesh00}, plus extensive
references in \cite{boon04}) have extended the eigen decomposition methods of
signal processing to applications in radio astronomy to capitalize on
these tools in the area of rfi mitigation. Here, we convey some of the
more intuitive conclusions to the astronomy audience.

\subsection{The covariance matrices}
\begin{figure*}
\noindent\includegraphics[width=39pc]{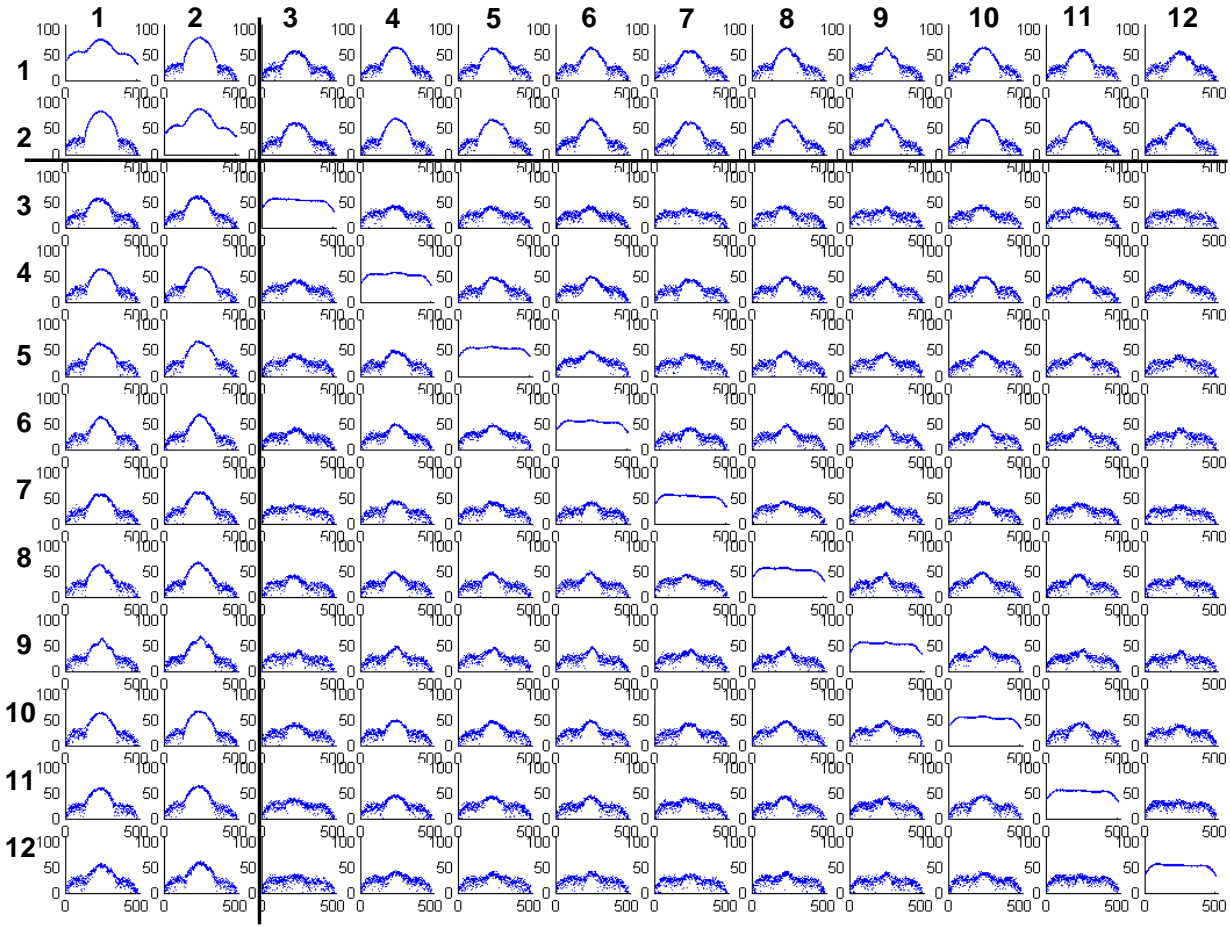}
 \caption{Examples of Power Spectra (diagonal elements) and Cross-Power Spectra
(off
diagonals) for a 12 signal cross correlation analysis. Each spectrum has 512
spectral channels with amplitudes in db. The upper left
$2{\times}2$ grouping contains the spectra and cross power spectra for the
reference channels. The lower right $10{\times}10$ group has the
spectra for the 5 antennas, with 2 polarizations each, for the ATCA telescopes.
The $2{\times}10$ and $10{\times}2$ groups to the lower left and upper right
respectively contain the cross correlation spectra that couple the reference
antennas to the ATCA signals.\label{cross_corr_spectra.fig}
}
\end{figure*}

To begin, Fig.~\ref{cross_corr_spectra.fig} displays a full set of cross power
spectra for a short observation with the Australia Telescope Compact Array --
with the spectra positioned according to their
positions in a correlation matrix.  This data set is ``srtca01'' from Bell et al.'s
baseband recordings (\cite{bell01}) taken specifically for the purpose of testing
rfi mitigation algorithms. The recordings include the time sequenced sampling of the
outputs from 5 antennas with two polarizations, plus two outputs from a dual polarized
reference antenna aimed at the source of interference, for a total of 12 input
data streams. The recorders captured 4 MHz bands with 4 bit precision. Correlation, to
obtain 512 spectral channels, and
integration were performed in software (\cite{kocz04}).
The rfi affects the central $\sim$150 spectral channels.

The total power and cross power spectra in Fig.~\ref{cross_corr_spectra.fig}
illustrate several points: The spectra formed from the reference data streams, 1 and
2, in the upper left show that the rfi is nearly 100\% linearly
polarized, as it appears
with equal strength in the total and cross power spectra. Further, the cross
power spectra for 1--2 show that the noise in the passband outside the $\sim$150
contaminated channels integrates down to a much lower level than in these spectral
ranges of the total power spectra for either 1--1 or 2--2; indeed, this is the
motivation for using the cross power spectrum instead of a total power spectrum
in the denominator of the correction term in Eqns.~\ref{totpwrcorr.eqn} and
\ref{crosspwrcorr.eqn}.  Similarly, the total power spectra along the diagonal
of the matrix in the lower right panels of Fig.~\ref{cross_corr_spectra.fig}
(involving inputs 3 through 12) scarcely show the presence
of the rfi in several cases, but the rfi clearly appears above the noise in
cross power spectra involving the same antennas.

\subsection{The eigen decomposition}

The eigen decomposition of a 12${\times}$12
covariance matrix solves for signals
that are in common among the 12 input data streams.
Since there are actually 512 covariance matrices (one for each of the
512 frequency channels), the decomposition must be performed 512 times. 
In matrix notation, one 12${\times}$12
covariance matrix \textbf{C},
\[
\textbf{C}=
\left(
\begin{array}{llll}
C_{1\:1} & C_{1\:2} & ... & C_{1\:12}\\
C_{2\:1} & C_{2\:2} & ... & C_{2\:12}\\
C_{3\:1} & C_{2\:2} & ... & C_{3\:12}\\
&... \\
C_{12\:1} & C_{12\:2} & ... & C_{12\:12}\end{array}
\right)
\]
decomposes to the product of three 12${\times}$12 matrices:
$\textbf{C}=\textbf{G}\textbf{P}\textbf{G}^{H}$,
where $\textbf{G}^{H}$ is the Hermitian conjugate of \textbf{G} and \textbf{P}
is a diagonal matrix containing the eigenvalues.
\[
\textbf{P}=
\left(
\begin{array}{ccccc}
P_1 & 0  & 0&... & 0\\
0 & P_2  & 0&... & 0\\
0 & 0  & P_3&... & 0\\
&... \\
0 & 0 & 0&... & P_{12}\end{array}
\right)
\]
It is convenient to think of the $P_i$ eigenvalues as powers.
Standard methods for
performing the eigen decomposition, such as singular value decomposition (SVD),
produce an ordered list of eigenvalues starting with the maximum. Should the
system have fewer than 12 signals present, the matrix would become singular, and
some number of eigenvalues would be zero.  This is not expected of a radio
astronomical system, where there are not only rfi signals and celestial sources
but also each input data stream brings its own noise signal.

The matrix \textbf{G} containing the eigenvectors
\[
\textbf{G}=
\left(
\begin{array}{llll}
g_{1\:1} & g_{1\:2}  & ... & g_{1\:12}\\
g_{2\:1} & g_{2\:2}  & ... & g_{2\:12}\\
g_{3\:1} & g_{2\:2}  & ... & g_{3\:12}\\
&... \\
g_{12\:1} & g_{12\:2} & ... & g_{12\:12}\end{array}
\right)
\]
can be visualized as a matrix of complex voltage gains $g_{ij}$,
giving the coupling of each signal (whose voltage is  $P_i^{1/2}$)
to the covariances. Thus, the amount of $P_i$ coupled to $C_{jk}$
is $g_{ji}g_{ki}^*P_i$. The top row of \textbf{G} specifies
the coupling of the 12 $P_i$ to the number 1 input, which is the
first reference antenna in the example here.

\begin{figure}
\noindent\includegraphics[width=20pc]{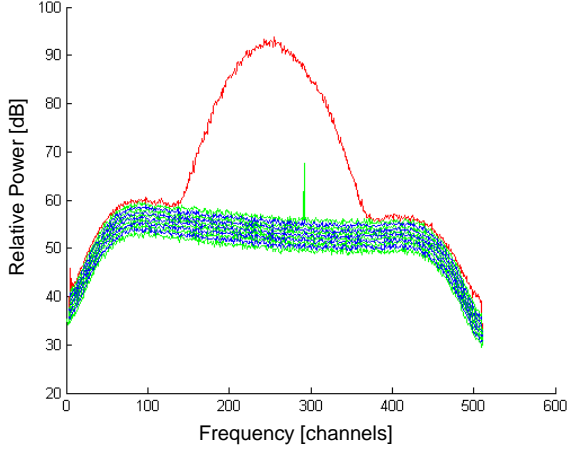}
\caption{Eigenvalue Spectra for $P_i(f)$ for $i$=1,12. Except for $P_1(f)$, the
spectra are very closely spaced, and the 11 parallel ``passbands'' are drawn in
alternating light and dark colors.
\label{eig_spectra.fig}
}
\end{figure}

Figure~\ref{eig_spectra.fig} shows the eigenvalue spectra
obtained from  512 eigen decompositions that were performed
one spectral channel at a time for the data in Fig.~\ref{cross_corr_spectra.fig}.
The calculation has succeeded in separating the rfi and concentrating 
its power in the
first eigenvalue $P_1(f)$. The other $P_i(f)$ for $i>1$ appear to represent normal
passband power levels expected from receiving systems dominated by internal noise.
The one exception is a narrow rfi spike in $P_2(f)$ around channel 290 that had
not been apparent in the spectra in Fig.~\ref{cross_corr_spectra.fig}.

\subsection{Special case: Array calibration}

An analogy can be made to a common procedure in the amplitude and phase
calibration of synthesis arrays. Typically, an observation is performed on
a bright, unresolved  radio continuum source, whose flux density $S_{\nu}$ 
is high enough that the source's power in the
receiving system will dominate over all other powers after a short integration
time. In the notation above, this condition means that there is one dominant
eigenvalue $P_1=S_{\nu}$. Under this condition, the \textbf{P} matrix degenerates
to 1$\times$1, and the eigenvector matrix \textbf{G} becomes an 12$\times$1
dimensional vector of complex gains.  In fact, the standard methods of 
array calibration solve for the ``antenna based complex gain factors'' (for
amplitude and phase).  These gains are a subset of the full eigenvector array,
but they enable the reconstruction of the covariance matrix for the case
with a single powerful source.

\[
\textbf{C}=
\left[
\begin{array}{l}
g_{1} \\
g_{2} \\
g_{3} \\
... \\
g_{12} \end{array}
\right]
\left[ S_{\nu} \right]
\left[ g_{1}^*\:
g_{2}^*\:
g_{3}^*\:
...\:
g_{12}^*
\right]
\]

\subsection{Nulling}
\label{nulling.section}

\begin{figure}
\noindent\includegraphics[width=20pc]{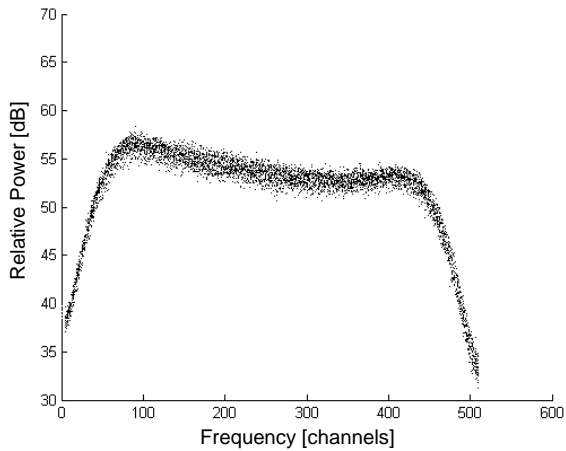}
\caption{Total power spectra for astronomy channels (3-12) after partial nulling
the dominant eigenvalue.
\label{tot_pwr_spectra_astro.fig}
}
\end{figure}

\begin{figure}
\noindent\includegraphics[width=20pc]{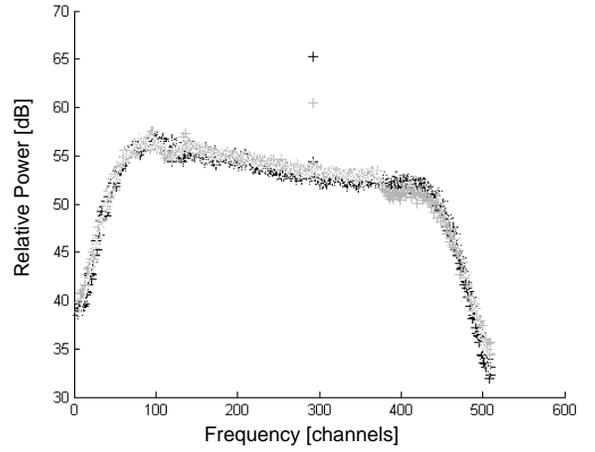}
\caption{Total power spectra for reference channels (1-2) after partial nulling
the dominant eigenvalue.
\label{tot_pwr_spectra_refs.fig}
}
\end{figure}

Having identified the strength and spectral behavior of the rfi through its
dominance in the eigen decomposition of Fig.~\ref{eig_spectra.fig},
one is tempted to null the rfi by
replacing $P_1$ by zero. However, this would be slightly too severe a partitioning
of the eigenvalue matrix, since this
would also remove the system noise contribution and perhaps distort a
weak celestial signal contributing noise power to the system. In these
illustrations reported here, 
we replace $P_1(f)$ by the average of the other $P_i(f)$ ($i$=2,12),
and then reconstruct the covariances $\textbf{C}(f)$ from \textbf{G} and the
modified \textbf{P} matrix. This produces
the total power spectra for the ten telescope data streams as shown
in Fig.~\ref{tot_pwr_spectra_astro.fig} and the two reference antenna
data streams in Fig.~\ref{tot_pwr_spectra_refs.fig}.

The partial nulling of the dominant eigen value has been effective in recovering
total power spectra for the telescope inputs (3--12) in which no remnant of the
rfi is visible. The partial nulling has also removed the rfi from the reference
inputs (1 and 2), allowing the second, narrow interferer from $P_2$
to be seen in the spectra for both. 
Apparently, this narrow interferer was only picked up by the
reference antennas and not by the ATCA telscopes themselves.

The process adopted here to partially null the eigenvalue and then recompute
an edited covariance matrix has the identical effect to the procedure in which the
unwanted portion of the eigenvalue \textbf{P} matrix is ``projected'' back 
into the covariance matrix and then subtracted (e.g. \cite{lesh00}).

\subsection{Directional information from the eigen decomposition}

\begin{figure}
\noindent\includegraphics[width=20pc]{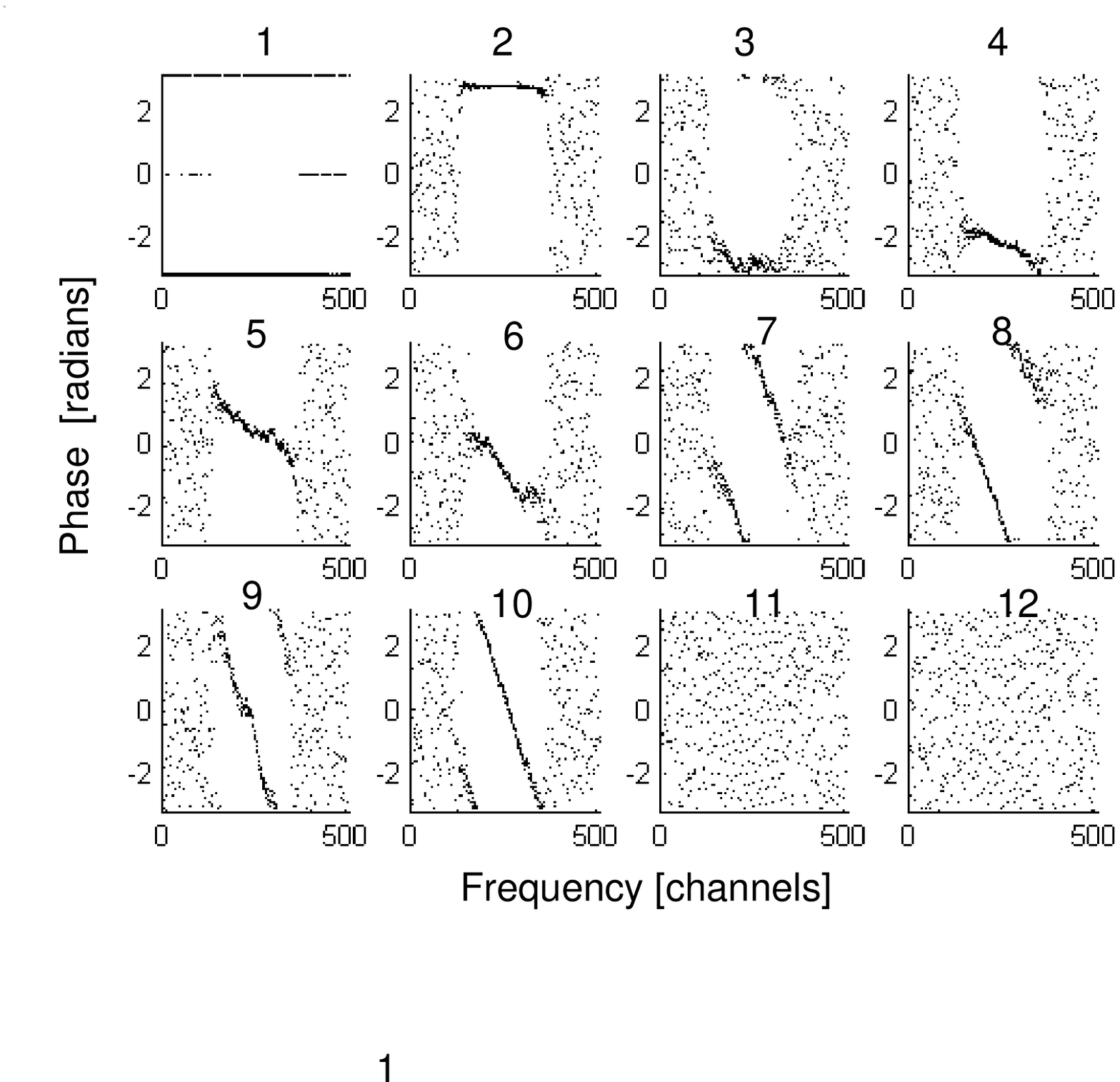}
\caption{Eigenvector phases for the dominant eigenvalue for all 12 data channels.
\label{eig_vec_phases.fig}
}
\end{figure}

In the worked example for 12 input data streams,
the eigen decomposition has identified two interfering sources
based on the mathematical interpretation of the convariance matrix, and it has also
enabled the nulling of the rfi power.  This is accomplished without knowledge
of location of the interferer, aside from having aimed the reference antennas;
in fact, basing the analysis on the 10 inputs from the ATCA dishes alone 
(and ignoring the reference antenna inputs) leads
to a similar result, albeit with a lower interference-to-noise ratio in the
eigenvalue spectra (\cite{kocz04}).

There is directional information contained in the phases of the complex eigenvector
coefficients. To illustrate this, Fig.~\ref{eig_vec_phases.fig} plots the phase
spectra for the $g_{1j}$ coefficients that couple $P_1$ (the dominant eigenvalue
containing the strong rfi power) to each of the array antennas. Outside the
central channels, the phases become random, implying that there is no strong
broadband signal in these spectral ranges 
to influence these spectral channels. In the center of the band
where the rfi is strong, the phases show the characteristic drift with frequency
that is expected for interferometer baselines of increasing length (and therefore
increasing delay) as indicated in Eqn.~\ref{phasewind.eqn}. The baseline length 
to the ATCA telescope number 6 was much longer that the other baselines between the
telescopes during this experiment, so the phases for data streams 11 and 12 that
were produced by telescope 6 wind very rapidly as a function of frequency
across the band and therefore are not detected by eye in Fig.~\ref{eig_vec_phases.fig}. 
Radio astronomers might expect to decode this phase information
to infer directional information about the origin of rfi signals -- a process
that would be straight forward if the positions of the elements of the array were
well known and the element gains were isotropic. In practice, the use of directional
information is hampered by the unknown and variable gains of the sidelobes through 
which the RFI enters the antenna (\cite{lesh00}).

\subsection{Subtleties and cautions}

The eigen decomposition approach shows promise for some applications in
rfi mitigation. There are some subtleties to its application that are covered
in greater depth in the series of papers by Leshem et al (e.g., \cite{lesh00}).
(1) Among these is the need to prewhiten the noise among the input channels in order
for the algorithm to perform well; this is a step that would require some care
in calibration of an array. (2) Further, there is the question touched upon in
section~\ref{nulling.section} where setting the dominant eigenvalue to zero
was attempted in order to null the rfi; there is some subtlety in how to choose
the right amount to remove and how much to leave behind, given that the signals
from celestial sources are likely to appear weakly in the eigenvalues.
(3) Astronomical covariance matrices generally carry signals from more sources than
the dimensionality of the matrix. While rfi might produce the strongest signals,
causing them to stand out in the eigen decomposition, there is also noise signal
 associated
with each of the telescope inputs, as well as with the signals from numerous weak
celestial radio sources. Once the number of signals exceeds the number of inputs
to the correlator, a complete recovery of the separate signals is no longer
possible; in effect, the number of unknowns exceeds the number equations
available for solution. Even the eigen values are then mixtures of signals, and
the process of nulling runs the risk of damaging the astronomical signals.

\section{Summary}

This paper gave an overview of a couple of methods for rfi cancellation that
show promise for application in radio astronomy observations. There remains
considerable challenge to achieving the dynamic range in rfi cancellation
likely to be required.

%
%

\begin{acknowledgments}
FHB is grateful to the staff of the Dominion Radio Astrophysical
Observatory in Penticton, B.C. for their fine job in hosting the RFI2004 Workshop.
The authors thank A.R. Thompson and D.A. Mitchell for valuable comments and 
discussion.
This research in rfi suppression was supported by the Australia Telescope National
Facility (ATNF) and an Australian Research Council grant.
\end{acknowledgments}

%
%
%
%
%
%
%
%


%
%
%

%
%

\end{article}

\end{document}